\documentclass[12pt]{article}
\usepackage{graphicx,epsfig}
\usepackage{color}

\def\be{\begin{equation}}
\def\ee{\end{equation}}
\def\ba{\begin{eqnarray}}
\def\ea{\end{eqnarray}}

\definecolor{verde}{rgb}{0.4,0.6,0}
\definecolor{azul}{rgb}{0.1,0.2,0.6}
\definecolor{rojo}{rgb}{0.8,0.1,0.1}

\begin{document}

\begin{center}

{\Large{\bf GRB 051221A and Tests of Lorentz Symmetry} }

\end{center}

\begin{center}

\bigskip

{\large Mar\'{\i}a Rodr\'{\i}guez Mart\'{\i}nez, Tsvi Piran and
Yonatan Oren}

\bigskip

Racah Institute of Physics,
The Hebrew University, \\
91904 Jerusalem, Israel. \\

\end{center}

\bigskip \bigskip

\begin{abstract}

  Various approaches to quantum gravity suggest the possibility of  violation
  of Lorentz symmetry at very high energies. In these cases we expect
  a modification at low energies of the dispersion relation of photons
  that contains extra powers of the momentum suppressed by a high
  energy scale. These terms break boost invariance and
  can be tested even at relatively low energies. We use
  the light curves of the very bright short Gamma-Ray Burst
  GRB 051221A and compare the arrival times of
  photons at different energies with the expected time delay due to
  a modified dispersion relation. As no time delay was observed, we
  set a lower bound of $0.0066 \, E_{pl} \sim 0.66 \cdot 10^{17}$ GeV on
  the scale of Lorentz invariance violation.

\end{abstract}


\setcounter{equation}{0}

\section{Introduction}

Various quantum gravity theories suggest that Lorentz symmetry is
broken or modified at very high energies
\cite{Amelino-Camelia:2002dx}. In most of these models the low
energy limit of the photon dispersion relation is deformed by an
addition of extra powers of the particle momentum. These
modifications are suppressed by a high energy scale, beyond which
Lorentz symmetry is broken. Among other implications, such terms
cause the photon speed to depend on the energy and to differ from
the classical speed of light, $c$ \cite{Coleman:1998ti}.

Amelino-Camelia et al. \cite{Amelino-Camelia:1997gz} proposed to
use the arrival times of photons from cosmological Gamma Ray
Bursts (GRBs) to set experimental bounds on the energy scale of
Lorentz invariance violation. The idea was pursued by several
groups which obtained observational bounds using ensembles of
bursts with known redshfits \cite{Ellis:2002in,Ellis:2005wr}, or
using a single very powerful and atypical burst GRB 021206
\cite{boggs-2004}. In this paper we use the very bright short
burst GRB 051221A to test Lorentz symmetry and set a new bound on
the possible scale of Lorentz symmetry violation.

\section{Lorentz Violation and Time Delay}

Following \cite{Amelino-Camelia-Piran:2001}, we consider deformations
of the dispersion relation of photons which break boost invariance but
keep rotational and translational symmetry:
\be
E^2 = {p^2  c^2 \over a^2} \left[1 +
\left( {p \, c \over  \xi E_{pl} \, a} \right)^n  \right]
 , \;\;\; n = 1, 2,... \;\;\;.
\label{mod_disp_rel}
\ee
The non-standard term is suppressed by the high energy scale, $\xi \,
E_{pl}$, where $\xi$ is a dimensionless parameter and $E_{pl}$ is the
Planck energy. This expression can be thought as a low energy
expansion (compared with $E_{pl}$) of a quantum-gravity
Hamiltonian. We will focus on $n=1$ and $n=2$ in the following.

Eq. \ref{mod_disp_rel} causes the speed of photons to depend on
their energy. Two photons emitted simultaneously with different
energies, arrive on Earth with a time delay $\Delta
t_{\mbox{{\scriptsize del}}}$. By comparing this time delay (or
the lack of it) with the time resolution of the observing detector
we can set bounds on the Lorentz violation parameter, $\xi$. The
time delay between two photons with energies $E_1$ and $E_2$
emitted simultaneously from a source at a redshift $z$
 is given by \cite{Martinez:2006ee}:
\be \Delta t_{\mbox{{\scriptsize del}}} \simeq {1+n \over 2 H_0 \,
  \xi^n} \, \left[ \left({E_2 \over  E_{pl} } \right)^n -
\left({E_1 \over  E_{pl}}\right)^n \right] \, { \sqrt{ \Omega_m
(1+z)^3 +
    \Omega_\Lambda}\over \sqrt{ \Omega_m + \Omega_\Lambda}} \int_0^z {
  (1+z)^n dz \over \sqrt{ \Omega_m (1+z)^3 + \Omega_\Lambda}} \; .
\label{time-delay}
\ee
We use in the following the ``standard" cosmological parameters $H_0$,
$\Omega_m$ and $\Omega_\Lambda$ \cite{Spergel:2003cb}.

\section{GRB 051221A}

GRB 051221A was a very powerful bright short burst located in a star
forming galaxy at a redshift of $z = 0.5465$ \cite{gcn-redshift}. It
was observed in the gamma ray regime by three different satellites:
Swift-BAT, Suzaku-WAM and Konus-Wind. Swift-BAT and Konus-Wind have
comparable instrumental time resolutions, 4 msec and 2 msec
respectively. The time resolution of Suzaku-WAM is, on the other hand,
significantly lower : $\sim$ 30 msec. Since we need the highest time
resolution for for our purposes \cite{Martinez:2006ee},
we only analyze the light curves of Swift-BAT and Konus-Wind which
provide the strongest bounds.

\subsection{The Swift-BAT data}

\begin{figure}
\begin{center}
\includegraphics[width=10cm]{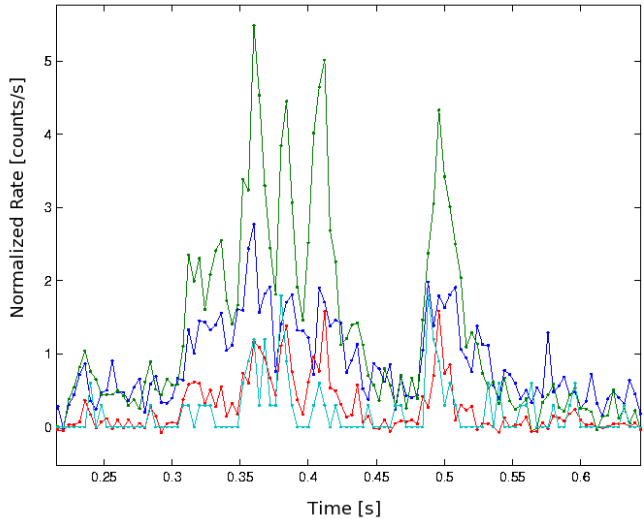}
\caption{From bottom to top, lights curves of GRB 051221A detected by
Swift-BAT in the energy bands 300 - 350 keV (light blue), 100 - 150 keV (red),
  15 - 35 keV (dark blue) and 50 - 150 keV (green). The background has been
substracted in all the bands except in 300 -350 keV, where the
noise hides the signal otherwise.
} \label{espectro-swift}
\end{center}
\end{figure}

The BAT light curve \cite{swift-data,swift-data2,swift-data3} is
depicted at Fig. \ref{espectro-swift}. It shows several bright short
peaks with a duration of $\sim 10-15$ msec each. Approximately a second
later a second smaller and softer peak followed lasting $\sim$ 3 sec.

The overall time resolution of a telescope is determined by the
intrinsic detector minimal time resolution, 4 msec for the Swift-BAT
data, and the photon arrival rate \cite{Martinez:2006ee}. For GRB
051221A, the resolution associated with the photon arrival rate in the
keV range is smaller than the instrumental resolution, 4msec. At the
higher energy band, (300-350) keV, the photon arrival rate becomes too
small and we need to switch to the unfiltered data. Overall, the
instrumental resolution is setting the limiting resolution.

To constrain the parameter $\xi$ we compare the light-curves covering
the (15, 35), (50, 150) keV, (100, 150) and (300, 350) keV energy
bands. We have measured the time positions of the four main peaks in
Fig.  \ref{espectro-swift} (measured in 4 msec time bins) :

\begin{center}
\begin{tabular}{|c|c|c|c|c|} \hline
 & \multicolumn{4}{c|}{Time position (s)} \\[3pt] \cline{2- 5}
& 1st peak & 2nd peak & 3rd peak & 4th peak \\[3pt] \hline
15 - 35 keV & 0.360 & 0.388 &0.408 & 0.496\\[3pt] \hline
50 - 150 keV & 0.360 & 0.388 &0.412 & 0.496\\[3pt] \hline
100 - 150 keV & 0.360 & 0.388 & 0.412 & 0.496\\[3pt] \hline
300 - 350 keV & 0.360  & 0.380 & 0.408 & 0.488\\[3pt] \hline
\end{tabular}
\end{center}
In the light curve in the 15 -35 keV band, the fourth peak is
composed of three sub-peaks. The time given in the table
corresponds to the middle sub-peak.  Notice that the first
sub-peak is simultaneous with the peak in the 300 -350 keV band.

At 4 msec time bin resolution we do not appreciate any time delay
between the different peaks in all three energy bands. This lack of
temporal shift is compared now with the theoretical times delays
expected from Eq. \ref{mod_disp_rel}:

\begin{center}
\begin{tabular}{|c|c|c|} \hline
& \multicolumn{2}{c|}{Time delay (s)} \\[3pt] \cline{2- 3}
& $n=1$ & $n=2$ \\[3pt] \hline
$\Delta t_{(15,100)\, \mbox{{\scriptsize keV}}}$ & $4.0 \cdot 10^{-6} / \xi$
& $5.6 \cdot 10^{-29} / \xi^2$ \\[3pt] \hline
$\Delta t_{(100,300)\, \mbox{{\scriptsize keV}}}$ & $9.4 \cdot 10^{-6} / \xi$
& $4.6 \cdot 10^{-28} / \xi^2$ \\[3pt] \hline
$\Delta t_{(15,300)\, \mbox{{\scriptsize keV}}}$ & $1.3 \cdot 10^{-5} / \xi$
& $5.2 \cdot 10^{-28} / \xi^2$ \\[3pt] \hline
\end{tabular}
\end{center}

As the photon flux decreases with energy, we consider in each energy
band the lowest energy as the relevant one for the calculation of
$\Delta t$. This conservative estimate lowers slightly the limit we
obtain on $\xi$. Since no time delay was observed, we can infer the
following bounds:

\begin{center}
\begin{tabular}{|c|c|c|} \hline
& $n=1$ & $n=2$ \\[3pt] \hline
$\Delta t_{(15,100)\, \mbox{{\scriptsize keV}}}$ & $\xi_1 >0.0010 $ &
$\xi_2 > 1.2 \cdot 10^{-13} $ \\[3pt] \hline
$\Delta t_{(100,300)\, \mbox{{\scriptsize keV}}}^*$ & $\xi_1 >
0.0023 $
&  $\xi_2 > 3.4 \cdot 10^{-13}$ \\[3pt] \hline
$\Delta t_{(15,300)\, \mbox{{\scriptsize keV}}}^*$ & $\xi_1 >
0.0033 $
&  $\xi_2 > 3.6 \cdot 10^{-13}$ \\[3pt] \hline
\end{tabular}
\end{center}
* The background signal has not been subtracted in the energy band
300-350 keV to keep an acceptable photon arrival rate.

\subsection{The Konus-Wind data}

\begin{figure}
\begin{center}
\includegraphics[width=13cm]{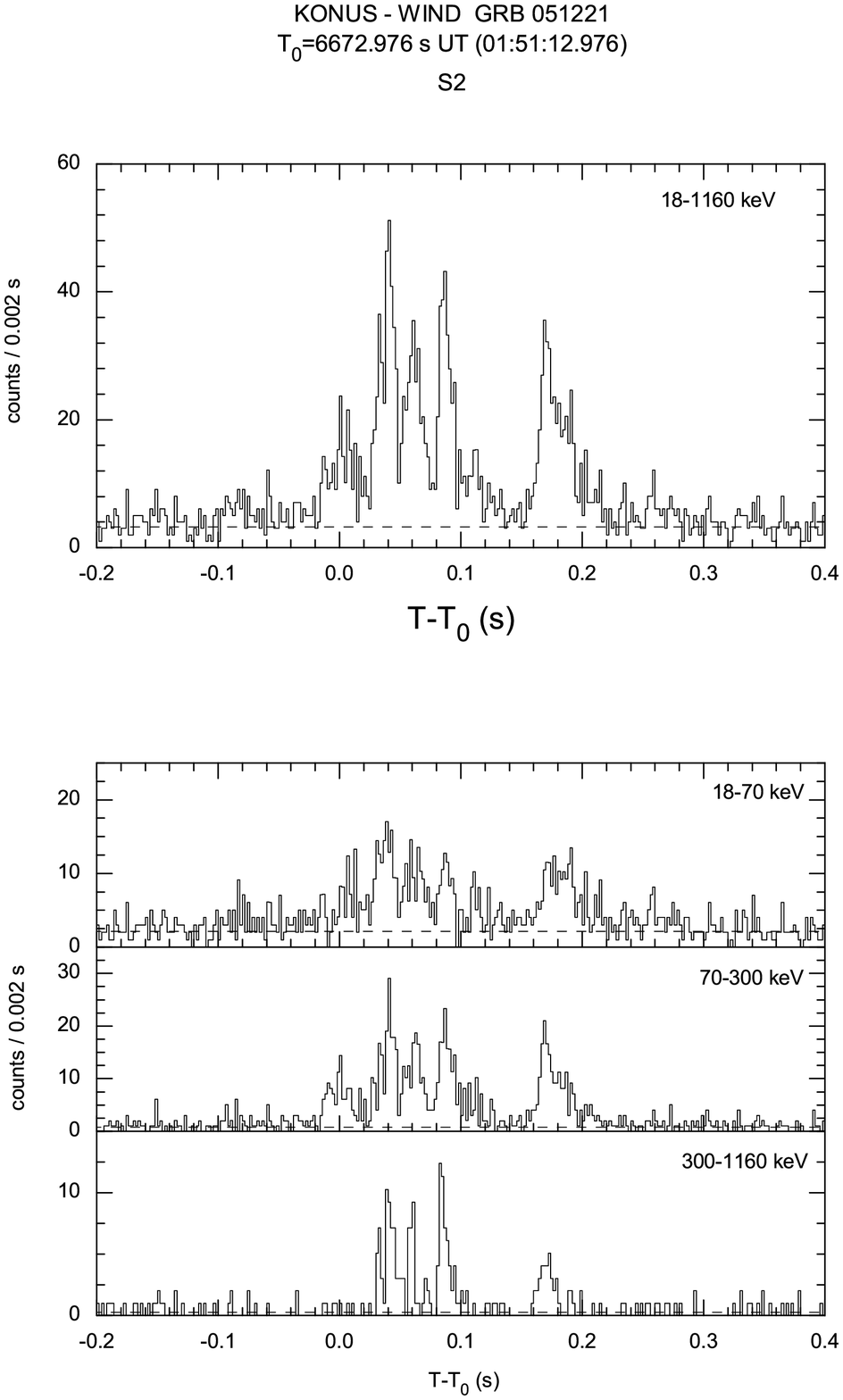}
\caption{Light curves of GRB 051221A as observed by Konus-Wind
\cite{konus-data}. From
http://www.ioffe.rssi.ru/LEA/GRBs/GRB051221\_T06672/ . }
\label{konus-lightcurve}
\end{center}
\end{figure}

The Konus-Wind light curve \cite{konus-data} consists of a soft weak
precursor and the main episode with five $\sim$ 15 msec peaks. The
first peak is substantially softer than the others (there is no
emission in the 380-1160 keV energy range). After 0.250 sec a weak
soft emission is marginally seen only in the 18-70 keV range up to
$\sim$ 1 sec (see Fig. \ref{konus-lightcurve}).

The light curve of Konus-Wind is given at 2 msec resolution,
corresponding to the intrinsic resolution of the telescope. As in
the Swift analysis, the resolution is dictated by the instrumental
resolution, even though the inverse photon rate is close to the
limiting resolution of the detector. At the high energy band there
are only a few photons per temporal bin. The light curve in the 18
- 1160 keV regime has been subdivided in three energy intervals :
18 - 70 keV, 70 - 300 keV and 300 - 1160 keV. The time positions of
the four main peaks are :


\begin{center}
\begin{tabular}{|c|c|c|c|c|} \hline
 & \multicolumn{4}{c|}{Time position (s)} \\[3pt] \cline{2- 5}
& 1st peak & 2nd peak & 3rd peak & 4th peak \\[3pt] \hline
18 - 70 keV & 0.038 & 0.064 & 0.086 & 0.16 \\[3pt] \hline
70 - 300 keV & 0.040 & 0.062 & 0.086 & 0.16 \\[3pt] \hline
300 - 1160 keV &0.038 & 0.060 & 0.082 & 0.16\\[3pt] \hline
\end{tabular}
\end{center}

The second peak in the energy band 18 - 70 keV is subdivided into two
peaks. We give the position of the second sub-peak where most of the
photons were detected.

The expected time delays are:

\begin{center}
\begin{tabular}{|c|c|c|}  \hline
& \multicolumn{2}{c|}{Time delay (s)} \\[3pt] \cline{2- 3}
& $n=1$ & $n=2$ \\[3pt] \hline
$\Delta t_{(18,70)\, \mbox{{\scriptsize keV}}}$ & $2.4 \cdot 10^{-6} / \xi$
& $2.6 \cdot 10^{-29} / \xi^2$ \\[3pt] \hline
$\Delta t_{(70,300)\, \mbox{{\scriptsize keV}}}$ & $1.0 \cdot 10^{-5} / \xi$
& $4.9 \cdot 10^{-28} / \xi^2$ \\[3pt] \hline
$\Delta t_{(18,300)\, \mbox{{\scriptsize keV}}}$ & $1.3 \cdot 10^{-5} / \xi$
& $5.2 \cdot 10^{-28} / \xi^2$ \\[3pt] \hline
\end{tabular}
\end{center}

Again, no time delays are observed and, hence, the following bounds
emerge

\begin{center}
\begin{tabular}{|c|c|c|} \hline
& $n=1$ & $n=2$ \\[3pt] \hline
$\Delta t_{(18,70)\, \mbox{{\scriptsize keV}}} $ & $\xi_1 >0.0012 $ &
$\xi_2 > 1.1 \cdot 10^{-13} $ \\[3pt] \hline
$\Delta t_{(70,300)\, \mbox{{\scriptsize keV}}}$  & $\xi_1 > 0.0054 $
&  $\xi_2 > 5.0 \cdot 10^{-13}$ \\[3pt] \hline
$\Delta  t_{(18,300)\, \mbox{{\scriptsize keV}}} $ & $\xi_1 > 0.0066  $
&  $\xi_2 > 5.1 \cdot 10^{-13}$ \\[3pt] \hline
\end{tabular}
\end{center}

These bounds are slightly better than the bounds obtained from the
Swift data. This improvement is simply due to the better intrinsic
time resolution, 2 msec of Konus-Wind versus 4 msec in Swift.

\section{Conclusions}

We set new bounds on the possible energy scale of Lorentz invariance
violation. To do so, we used the simultaneity of peak emission times
in the light curves in different energy bands observed by Swift-BAT
and Konus-Wind for GRB 051221A. The order of magnitude of the bounds
depends on how the photon dispersion relation is deformed. In the
models where a cubic term in the momentum is added ($n=1$) the best
bound we obtained is 0.0066 $E_{pl}$. When a quartic term is added
($n=2$) the effect is much smaller and the corresponding best bound is
$5.1 \cdot 10^{-13} E_{pl}$.

The present work complements previous studies which found bounds of
similar order of magnitude. Ellis et al.
\cite{Ellis:2002in,Ellis:2005wr} performed a statistical analysis in
an ensemble of bursts with known redshifts (to correct for intrinsic
time delays), finding the bounds $\xi_1 > 5.6 \cdot 10^{-4} $ and
$\xi_2 > 2.4 \cdot 10^{-13}$ at a 95\% of confidence level. In
general, intrinsic delays render difficult to constrain Lorentz
violations using a single burst, but we can still extract lower limits
if there is simultaneity in the peak emissions. In our study, since no
time delays are observed in the light curves with 2 msec bins, we can
assume that, at that resolution, all photons were emitted
simultaneously.  This is, of course, an assumption. We cannot rule out
the possibility in which intrinsic spectral temporal shifts are
exactly compensated by the time delays produced by the modified
dispersion relation, Eq.  \ref{mod_disp_rel}. While, in principle,
this is possible, we believe it to be extremely unlikely.

Boggs et al. \cite{boggs-2004} considered an unique extremely bright
burst GRB 021206 observed by RHESSI. This burst has an almost flat
spectrum in the band 1 - 17 MeV, which allowed estimates of the time
delay up to 17 MeV. The redshift of GRB 021206 is not known as no host
galaxy was detected. Assuming a redshift of $z \simeq 0.3$ and using
the simultaneity of the peaks at different energies, Boggs et al.,
\cite{boggs-2004} obtained the bounds $\xi_1 > 0.015 $ and $\xi_2 >
4.5 \cdot 10^{-12}$ for $n=1,2$.  These lower bounds are higher than
ours by factors of 2.5 and 9 respectively.  However, the redshift of
this burst was not measured, but only estimated from the spectral and
temporal properties of the burst.  From this point of view, the present
estimate, which is based on a burst with a known redshift, is more
robust.

\section*{Acknowledgments}

We thank Kim Page and the UK Swift Science Data Centre for kindly
providing us with the Swift-BAT light curves of GRB 051221A. This
research was supported by the EU-RTN ``GRBs - Enigma and a Tool'', by
an US-Israel BSF grant and by the Schwarzmann university chair (TP).

\end{document}